%% file: cen_acf.tex
\documentclass[useAMS,usenatbib]{mn2e}

\include{defn}
\usepackage{graphicx}
\usepackage{amsmath}

\title[Searching for the missing iron in the core of the Centaurus cluster]{Searching for the missing iron mass in the core of the Centaurus cluster}
\author[E. K. Panagoulia, A. C. Fabian \& J. S. Sanders]{E. K. Panagoulia$^{1}$\thanks{E-mail:
caillean@ast.cam.ac.uk}, A. C. Fabian$^{1}$ and J. S. Sanders$^{2}$\\
$^{1}$Institute of Astronomy, Madingley Road, Cambridge CB3 0HA\\
$^{2}$Max-Planck-Institute f\"{u}r extraterrestrische Physik, 85748, Garching, Germany.}
\begin{document}

\date{Accepted . Received ; in original form }
\pagerange{\pageref{firstpage}--\pageref{lastpage}} \pubyear{2002}

\maketitle

\label{firstpage}

\begin{abstract}
We re-analyse a combined 198 ks {\it Chandra} observation of NGC~4696,
the brightest galaxy of the Centaurus cluster. We extract
temperature and metallicity profiles from the data, and we confirm the
presence of a sharp drop in iron abundance, from $\sim$1.8 Z$\odot$ to
$\sim$0.4 Z$\odot$, within the central 5 kpc of the cluster. We
estimate that this central abundance drop corresponds to a total
``missing'' iron mass of 1.4$\times$10$^{6}$ M$_{\odot}$. We propose
that part of this missing iron is locked up in cool ($\sim$19 K),
far-IR emitting dust, as found by {\it Spitzer} and {\it Herschel}
observations. This can occur if the iron injected by stellar mass loss
in the central region is in grains, which remain in that form as the 
injected dusty cold  gas  mixes and joins the cold
dusty filamentary nebula observed within the same region. The bubbling
feedback process observed in the cluster core then drags filaments 
outward and dumps them at 10--20\,kpc radius, where the metallicity is 
high.  
   \\
\end{abstract}


\section{Introduction}
Galaxy clusters are the brightest extended X-ray sources in the Universe. The intracluster medium (ICM) is observed in the X-ray through the emission of thermal bremsstrahlung radiation and line emission, as it is gravitationally compressed and heated to temperatures of 10$^{7}$--10$^{8}$ K. Spectroscopic studies of the ICM have shown that it is enriched with metals expelled from the cluster galaxies, while an average background metallicity of $\sim$0.3 solar has also been observed \citep[e.g.][]{Edge91}. By comparing the ICM metallicity profiles with those produced by supernova simulations, it was proposed that this background enrichment is due to supernovae, which occured early in the life of the cluster \citep{Finoguenov00, Loewenstein06, dePlaa07}.  

Previous studies of galaxy clusters have allowed us to distinguish two different types of clusters, on the basis of their surface brightness profile. The first type are the clusters which show a sharp surface brighness peak towards the centre; these are cool-core (CC) clusters. Non cool-core clusters do not show such a surface brightness peak. Cool-core clusters display little recent merger activity, and are believed to be dynamically relaxed. The ICM cooling time in these CC clusters is often much shorter than the age of the cluster itself. Hence, according to the cooling flow model \citep[for a detailed description, see][]{Fabian94}, large amounts of cold gas and star formation are expected to be found at the cores of these clusters. However, detailed X-ray observations of these clusters have shown that the amount of gas with a temperature 2--3 times lower than the ambient gas, is much smaller than expected from the cooling flow model \citep{McNamara07, McNamara12, Peterson06}. In addition, the cooling flow model severely overpredicts the star formation rates in clusters \citep[see e.g.][]{Nulsen87}. This clearly indicates that a source of heating is needed, that will stop the gas from cooling and condensing onto the central cluster galaxy. The main driving mechanism behind this heating is thought to be AGN activity, a theory which has been supported by a wealth of studies on galaxy clusters \citep[e.g.][]{Birzan04, Dunn06, Rafferty06, McNamara12, Fabian12}. 

Although AGN activity and other heating mechanisms prevent the ICM from cooling catastrophically, recent observations of brightest cluster galaxies (BCGs) have indicated that the ICM is cooling, albeit at a suppressed rate. In fact, these BCGs seem to possess significant amounts of cold gas, a small part of which is in the process of star formation \citep{Johnstone87, Allen95, Odea08, Odea10}. In addition, many CC clusters exhibit optical line filaments, e.g. Abell~1795 (Crawford, Hatch et~al. 2005). It is therefore clear that gas exists in several different phases across a wide temperature range, from the cluster virial temperature (10$^{7}$--10$^{8}$ K), down to the temperature of molecular clouds ($\sim$10 K), where most of the cold mass in clusters resides \citep{Edge01}. 

\begin{figure*}
  \includegraphics[trim = 3.0cm 2.0cm 2.0cm 1.2cm, clip, height=7.0cm, width=8.5cm]{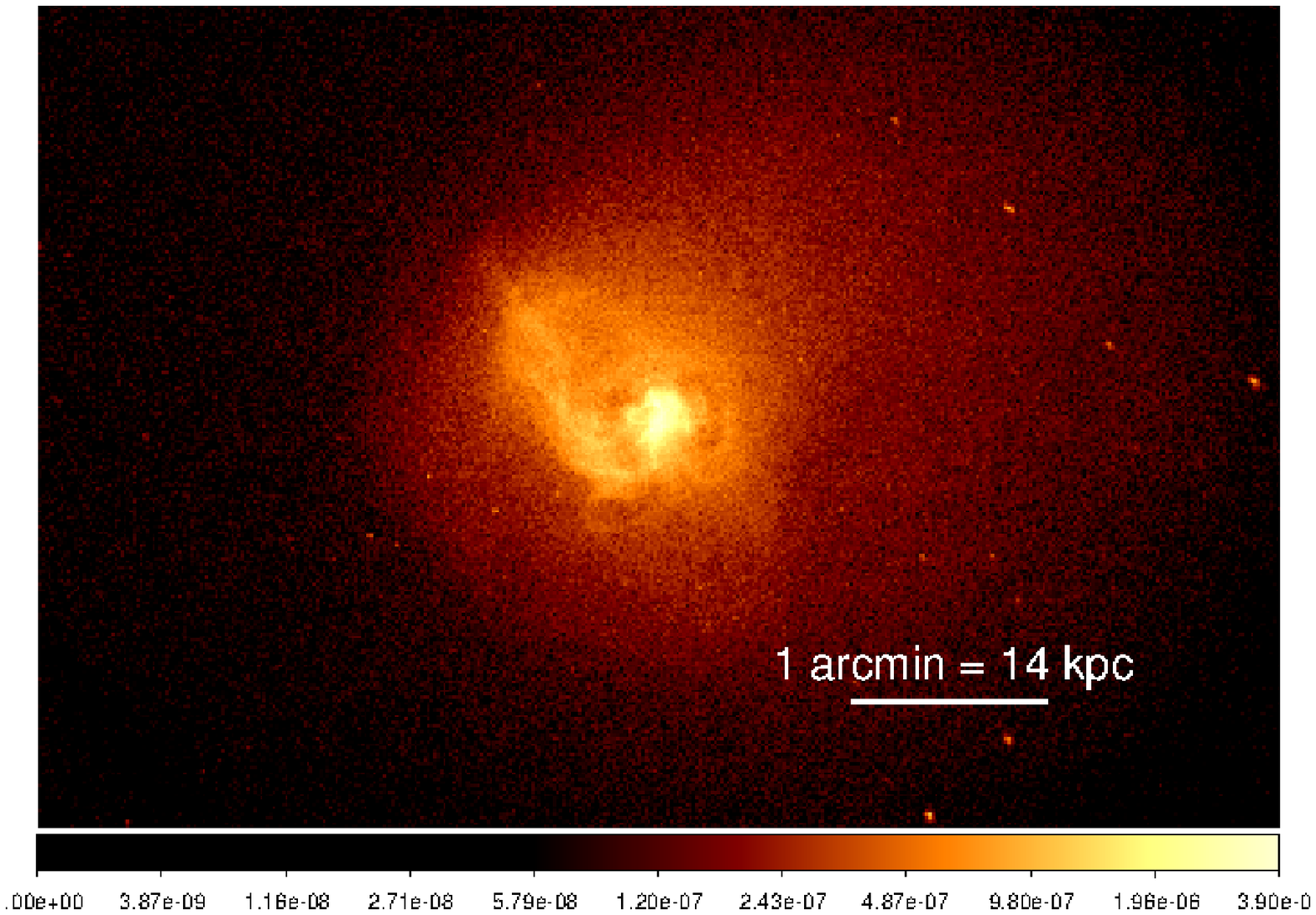}
  \hspace{0.3cm}
  \includegraphics[trim= 2.7cm 2.0cm 0.0cm 0.0cm, clip, height=7.2cm, width=8.5cm]{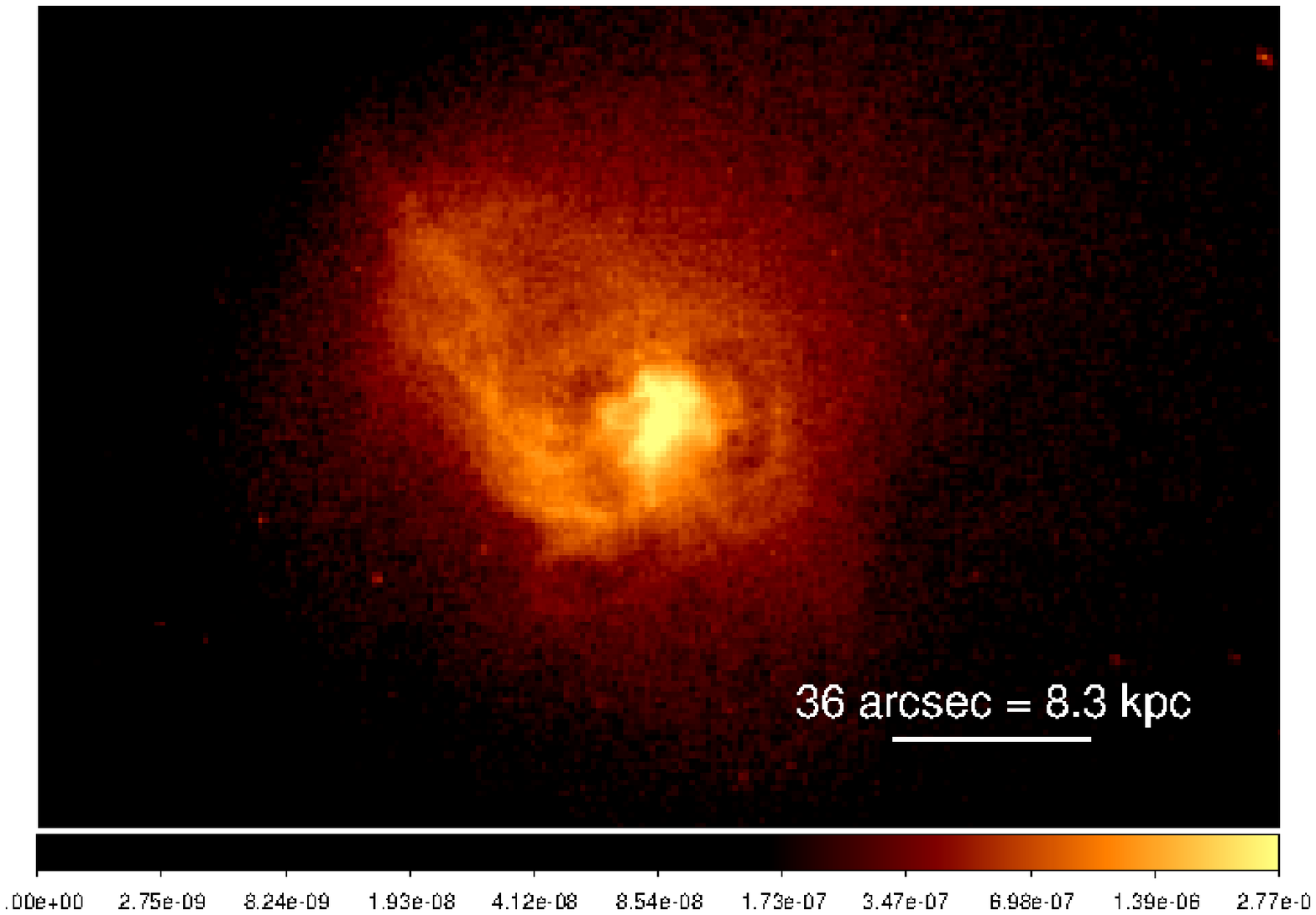}
  \caption{{\it Left}: 0.5--7.0 keV composite image of NGC~4696. Two surface brightness edges, one to the west and one closer in to the east of the cluster, are visible. A plume of gas is seen extending to the north, and the two X-ray cavities are clearly visible in this image. North is towards the top of the image, and west is to the left. {\it Right}: 0.5--7.0 keV composite image of the core of NGC~4696.}
  \label{fig:expcorr}
\end{figure*}

In this paper, we study NGC~4696, the BCG of the Centaurus galaxy cluster. NGC~4696 has a redshift of 0.0114 \citep{Struble99}, and has been studied extensively, in all wavelengths, because of its proximity \citep{Fabian82, Lucey86, deJong90, Allen94, Sanders02, Sanders06a, Sanders08, Crawford05b, Johnstone07, Canning11b, Mittal11}. Centaurus is a textbook example of a CC cluster, hosting a relatively weak cooling flow of a few tens of M$\odot$ yr$^{-1}$ \citep[e.g.][]{Allen01}, and having a central cooling time of $<$0.5 Gyr. A cool-temperature component, at $\sim$ 1 keV, in addition to a hotter $\sim$4 keV component, was first detected by {\it ASCA} \citep{Fukazawa94} and {\it ROSAT} \citep{Allen94}. The two-temperature model for the ICM was later on confirmed by \cite{Allen01} and \cite{Molendi02}. Deep {\it Chandra} images have revealed a bright core, as well as a plume of gas at the centre of the cluster, extending to the North \citep{Sanders02}. \cite{Sanders02} also showed that the temperature structure of Centaurus is bimodal, with a lower temperature towards the cluster core, and asymmetric in the east-west direction. Both {\it Chandra} and {\it XMM-Newton} observations have revealed that the Centaurus cluster has a complex metallicity structure, with steep abundance peaks towards the centre \citep{Sanders02, Sanders08}. Using data from the Reflection Grating Spectrometer (RGS) on board {\it XMM-Newton}, \cite{Sanders08} find that cooling through X-ray emission below 0.8 keV falls to $<$4 M$\odot$ yr$^{-1}$. \cite{Sanders06a} reported enrichment of the ICM gas by a combination of Type Ia and Type II supernovae. Using {\it Chandra} data, the same authors report a drop in the iron abundance in the central 10 kpc of the Centaurus cluster. 

In the optical, Centaurus displays a series of bright line-emitting filaments \citep{Fabian82, Crawford05b}, which match the X-ray structure of the cluster, as well as a deep dust lane \citep{Crawford05b}. Using {\it Herschel}, \cite{Mittal11} detected cold ($\sim$ 19 K) and warm ($\sim$46 K) dust emission, with the cold dust component being 400 times more massive than the warm dust component. The same authors also calculated a star formation rate (SFR) of 0.13 M$_{\odot}$ yr$^{-1}$. {\it Spitzer} observations detected emission from NGC~4696 in the near-, mid- and far-infrared, which originates from polycyclic aromatic hydrocarbons (PAHs), circumstellar dust emission and interstellar dust, respectively \citep{Kaneda05, Kaneda07}. 

Member galaxies of  CC clusters are expected to enrich their surrounding gas with metals from supernovae and galactic winds. However, the Centaurus cluster is just one of a number of clusters exhibiting a central abundance drop. Other clusters with central abundance dips include Abell~2199 \citep{Johnstone02}, Abell~1644 \citep{Kirkpatrick09} and the Perseus cluster \citep{Schmidt02, Churazov03, Sanders04}. Central abundance dips are also not limited to clusters. \cite{Rafferty13} report a central abundance drop in the centre of HCG~62, from $\sim$0.8 Z$\odot$ at a radius of 15'' to 0.3 Z$\odot$ in the central 5''. \cite{Rasmussen07} derived iron abundance profiles for 15 Hickson compact groups, showing a dip in the iron abundance in the very centre of most of the groups with a sufficiently resolved core. One of the suggested causes of these abundance dips is resonance scattering \citep{Gilfanov87, Sanders06b}. According to this theory, if a galaxy cluster is sufficiently optically thick at the energies of strong resonance lines, radiation that would have otherwise travelled along the line of sight of an observer, is now scattered away from that line of sight. If multiple scattering occurs, photons may even be absorbed by cold gas. As a result, the abundance in the centre of the cluster could be underestimated, while the abundance towards the outer regions could be overestimated. However, \cite{Sanders06b} found that resonance scattering does not explain the central abundance dip in Centaurus, although they indicate that internal absorption has a significant effect in the innermost 40 kpc of the cluster.  

In this paper, we adopt the cosmology H$_{0}$ = 70 km s$^{-1}$ Mpc$^{-1}$, $\Omega_{\rm m}$=0.27 and $\Omega_{\rm \Lambda}$= 0.73. All abundances presented in this paper are relative to solar, as defined in \cite{Anders89}. 

\section{Observations and data preparation}
In our analysis, we used the archival {\it Chandra} datasets with observation IDs 4954, 4955, 504 and 5310. In all these observations, the Centaurus cluster was placed on the back-illuminated S3 chip of the ACIS-S detector. The datasets were reprocessed using the {\sc ciao acis\_reprocess\_events} pipeline, and the appropriate charge transfer inefficiency correction files and gain files were applied to calibrate the data. Each dataset was individually examined, in order to excise periods of background flaring. To achieve this, 2.5--7.0 keV lightcurves were extracted from the back-illuminated S1 chip of the ACIS-S detector. These were then visually examined for periods of background flaring, which were excluded from any subsequent analysis. This process yielded a total exposure time of 198 ks. After the lightcurve filtering, each dataset was reprojected onto dataset 4954. 

Blank-sky observations were selected and adjusted to match the individual observations. To account for the increase in background with time in the background observations, their exposure times were altered so that their count rate in the 9.0--12.0 keV energy band matched that of the corresponding cluster observation. The background datasets were then also reprojected onto dataset 4954. For the creation of the total background image, the exposure time of each reprojected background image was weighted by the ratio of the cluster observation exposure time over the total exposure time (198 ks). The individual weighted, reprojected background images were then added together to create a total background image. 

A 0.5--7.0 keV background-subtracted, exposure-corrected composite image of the Centaurus cluster was created and is shown in Fig. \ref{fig:expcorr}. The {\sc ciao} {\sc wavdetect} routine was used to identify point sources in this image. The point sources were then visually verified and excluded from all subsequent spectral analysis.

\section{Spectral analysis and results}
\subsection{Analysis}
The main aim of our analysis is to study the radial profiles of the cluster temperature, metallicity and enclosed mass. To this end, we extracted spectra from concentric annular regions, centred on the core of the cluster, which we define as the centre of the X-ray emission peak (RA 12h48'48.8'', Dec -41$^{\circ}$18'45.0''). The annular regions were created in such a way that they all contained the same number of counts, i.e. they all had the same signal-to-noise ratio. The total background-subtracted, but not exposure-corrected, cluster image was used to define the annular regions, to ensure that the extracted spectra represented the same region in all the datasets. 

We created 22 annuli from the total background-subtracted cluster image, each one containing 115000$\pm$5000 counts in the 0.5--7.0 keV band, corresponding to a signal-to-noise ratio of $\sim$340. These annuli cover 0.11--70.05 kpc in radius from the defined cluster centre. We extracted foreground spectra, and the corresponding background spectra, for each annulus from each of the 4 datasets. Corresponding ancillary region files (ARFs) and response matrix files (RMFs) were created using {\sc mkwarf} and {\sc mkacisrmf}, respectively. To account for projection effects, the spectra were deprojected using the {\sc dsdeproj} routine, as described in \cite{Sanders07} and \cite{Russell08}. {\sc dsdeproj} is a model-independent deprojection method, which, under the only assumption of spherical symmetry, removes projected emission in a string of shells. After deprojection, the spectra were binned to have a minimum of 25 counts per spectral bin. The joint spectral fits were performed in {\sc Xspec} \citep{Arnaud96} v12.7.1b in the 0.5--7.0 keV energy band. The model used for the cluster emission was an absorbed, optically thin, single-temperature thermal plasma model, {\sc wabs*vapec}. The {\sc wabs} component allows for photoelectric asborption between the cluster and Earth, and the {\sc vapec} term is used to model the thermal emission from the cluster. 

The available free parameters in our {\sc wabs*vapec} model are: the column density nH, the temperature, the abundance of He, C, N, O, Ne, Mg, Al, Si, S, Ar, Ca, Fe and Ni, the redshift of the source and the normalisation of the thermal {\sc vapec} component. We used a fixed column density of 8.31$\times$10$^{20}$ cm$^{-2}$, and a fixed source redshift of 0.0114. We also fixed the He abundance to solar. For each set of four spectra corresponding to the same annular region, we allowed the temperature and iron abundance to vary independently, though their values for the four spectra were tied together. The parameter values of the rest of the abundances were tied to that of iron in the same annulus. To account for the differences in the exposures of the four datasets, the normalisation of the {\sc vapec} component was left free to vary, for each of the spectra. We note that, although varying independently, the normalisations for the {\sc vapec} component for the spectra corresponding to the same annulus, had very similar values in the resulting spectral fit. 

\begin{figure}
  \includegraphics[trim=0cm 14cm 5cm 0cm, clip, width=8.5cm, height=9.5cm]{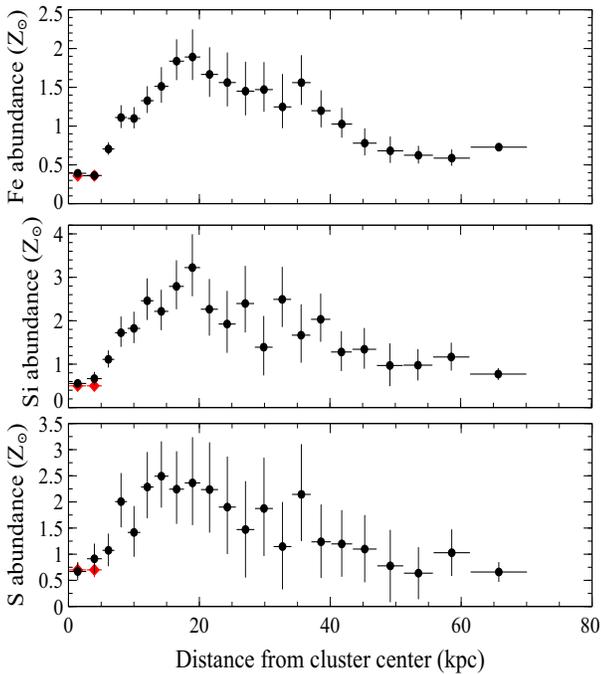}
  \caption{Deprojected iron (top panel), silicon (middle panel) and sulphur (bottom panel) abundance profiles for NGC~4696. The red points denote the best-fit abundances from the two-temperature fit of the inner two spectral bins. Errors are at the 90\% confidence level.}
  \label{fig:abund}
\end{figure}

\begin{figure*}
  \includegraphics[trim=0cm 0cm 0cm 0cm, clip, height=8.4cm, width=8.4cm]{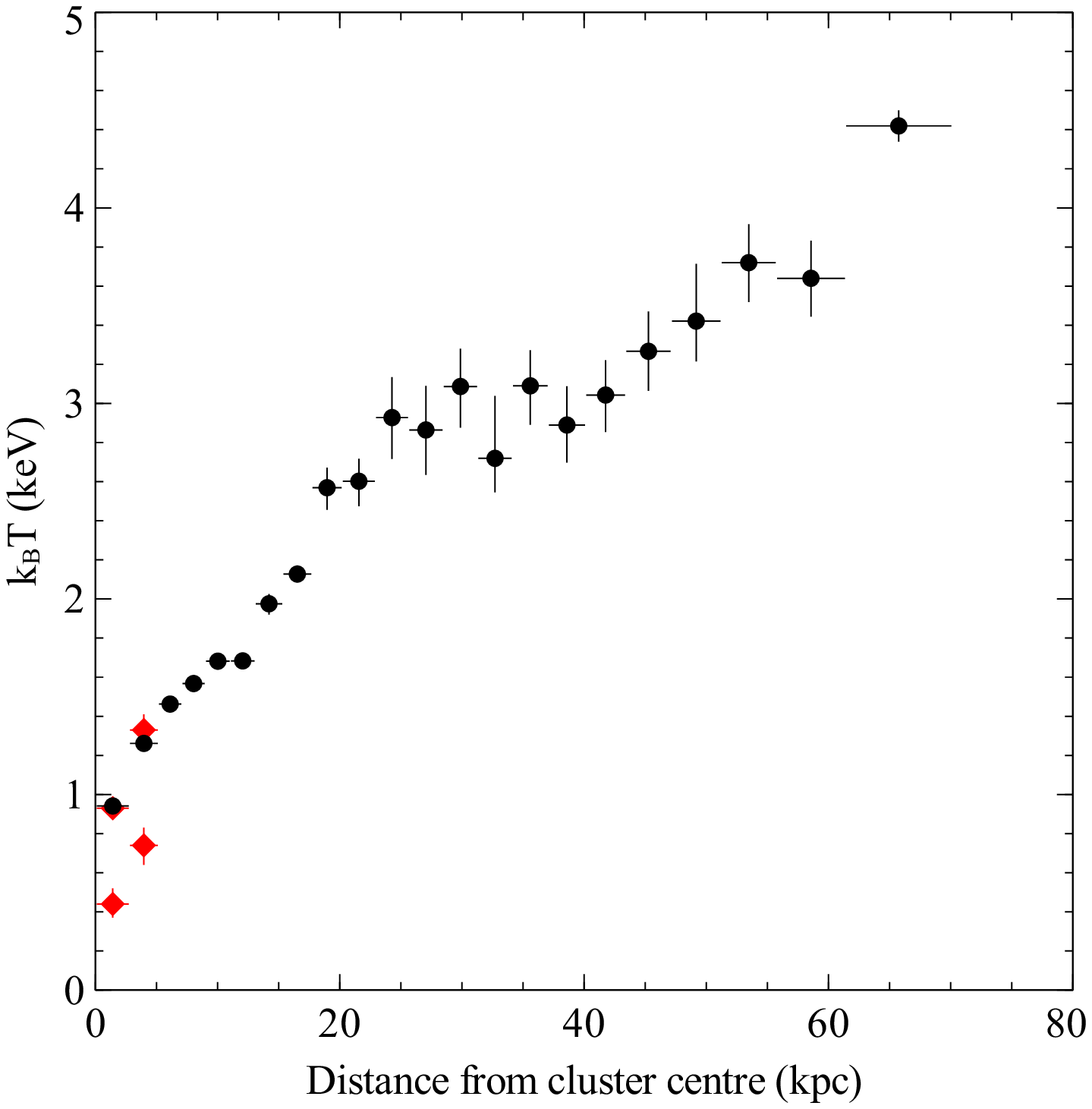}
  \includegraphics[trim= 0cm 0cm 0cm 0cm, clip, height=8.4cm, width=8.4cm]{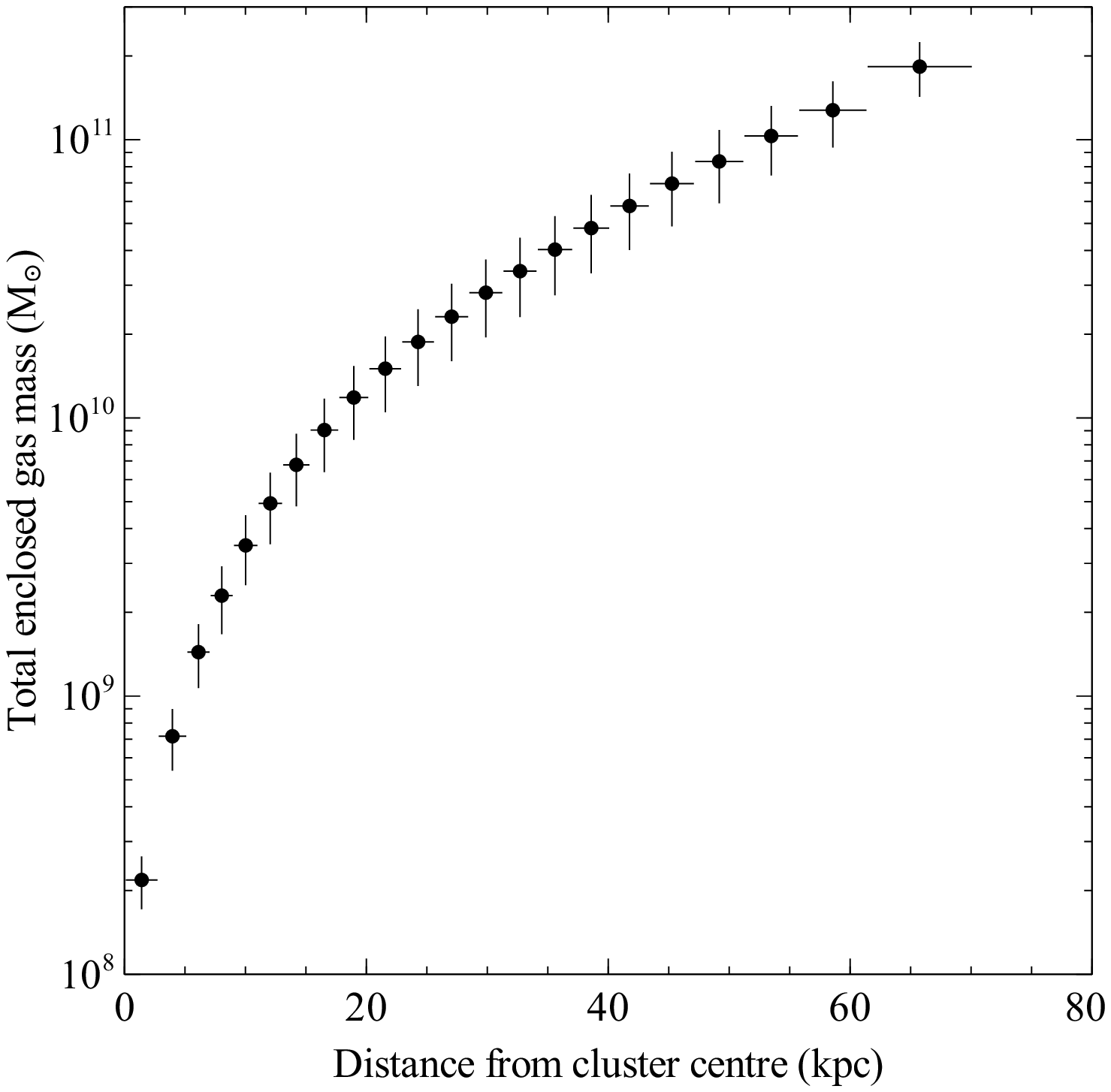}
  \caption{{\it Left}: Deprojected temperature profile for NGC~4696. The red points denote the best-fit temperatures from the two-temperature fit of the inner two spectral bins. {\it Right}: Total enclosed gas mass profile. Errors in all plots are at the 90\% confidence level.}
  \label{fig:temp}
\end{figure*}

\subsection{Results} 
The spectral fit we performed gave a statistically acceptable fit to the data. The resulting deprojected radial temperature and iron, silicon and sulphur abundance profiles are shown in the left panel of Fig. \ref{fig:temp} and Fig. \ref{fig:abund}, respectively. The total enclosed mass profile is shown in the right panel of Fig. \ref{fig:temp}. The temperature profile shows some variability in the 20--45 kpc region. The behaviour in this region is similar to that reported in \cite{Sanders02}, where they fit their data with absorbed two-temperature models ({\sc wabs*(apec+apec)}). The same authors attributed this behaviour to the asymmetry between the temperature profiles in the east and west of the cluster. We note that there are visible surface brightness discontinuities (Fig. \ref{fig:expcorr}) at a distance of 16'' to the east and 44.9'' west of the cluster centre, consistent with the temperature jumps reported in Fig. 9 of \cite{Sanders02} and in \cite{Fabian05}. As we are primarily interested in the iron abundance profile of NGC~4696, we do not investigate the temperature variability in the 20--45 kpc region any further in this paper. 

There is an obvious drop in the iron abundance profile within the central 7 kpc of the cluster, as shown in the top panel of Fig. \ref{fig:abund}. This drop is particularly steep in the innermost 5 kpc. The abundance profile peaks at 1.8 Z$_{\odot}$ at around 20 kpc, and then starts to decline relatively smoothly to 0.6 Z$_{\odot}$ at $\sim$ 60 kpc. Since we deprojected our spectra before performing the spectral fits, it is unlikely that this central abundance drop is caused by projection effects, and must therefore be real. 

\subsection{Two-temperature fit for the cluster centre}
Previous studies of the Centaurus cluster, using both {\it XMM-Newton} RGS data \citep{Sanders08} and {\it Chandra} data \citep{Sanders02}, showed that the temperature profile in its innermost regions is bimodal. Based on these findings, to calculate the missing iron mass within 5 kpc, we performed a joint spectral fit of the two innermost spectral bins (which correspond to the inner 5 kpc), using a two-temperature absorbed thermal model, {\sc wabs*(vapec+vapec)}. In this case, N$_{{\rm H}}$ was allowed to vary freely, as were the abundances of O, Ne, Mg, Si, S, Ca, Fe and Ni. 
The rest of the abundances in the model, namely C, N, and Al, were frozen at 0.3 Z$_{\odot}$, as the energy range of {\it Chandra} makes the spectra insensitive to C and N. All abundances were tied together for the two temperature components. For both spectral bins, the temperatures and normalisations for each {\sc vapec} component were left as free parameters, and all the respective abundances were tied. As expected, fitting a two-temperature model to the data, rather than a single-temperature model, significantly improves the quality of the fit. Using the same free parameters, the $\chi^{2}$/dof decreases from 1384.32/1190 for the single-temperature fit, to 1207.87/1187 for the two-temperature fit. The best-fit values for the parameters of the two-temperature spectral fit are listed in Table \ref{tab:results}. The best-fit temperatures and abundances are shown as red points in the left panel of Fig. \ref{fig:temp} and Fig. {\ref{fig:abund}, respectively. 

\renewcommand{\arraystretch}{1.2}
\begin{table}
  \caption{Best-fit parameter values for joint fit of inner two spectral bins. The errors shown are at the 90\% confidence level, and the abundances are relative to solar.}\label{tab:results}
\begin{center}
  \scalebox{1.0}{
    \begin{tabular}{ccc}\hline
      \multicolumn{2}{c}{Parameter value} & Best-fit value \\
      \hline
      1st spectral bin & k$_{\rm{B}}$T (keV) & 0.44$^{+0.09}_{-0.06}$  \\
        & Norm & (3.52$^{+1.28}_{-0.77}$)$\times$10$^{-4}$   \\
         &  k$_{\rm{B}}$T (keV) & 0.93$\pm$0.02 \\
         & Norm & (1.10$_{-0.10}^{+0.12}$)$\times$10$^{-3}$ \\
         \hline
        2nd spectral bin & k$_{\rm{B}}$T (keV) & 0.77$^{+0.06}_{-0.12}$ \\
         & Norm & (2.64$^{+0.67}_{-0.82}$)$\times$10$^{-4}$\\
          & k$_{\rm{B}}$T (keV) & 1.35$^{+0.08}_{-0.06}$ \\
         & Norm & (8.56$_{-0.95}^{+1.37}$)$\times$10$^{-4}$ \\
         \hline
         O & & 0.12$^{+0.08}_{-0.07}$ \\
         Ne & & 0.38$\pm$0.22 \\
         Mg & & 0.46$_{-0.11}^{+0.13}$ \\
         Si & & 0.54$^{+0.08}_{-0.07}$ \\
         S & & 0.76$^{+0.15}_{-0.14}$ \\
         Ca, Ar & & 1.24$^{+0.58}_{-0.56}$\\
         Fe & & 0.37$^{+0.06}_{-0.05}$ \\
         Ni & & 1.60$^{+0.54}_{-0.50}$ \\
         N$_{\rm{H}}$ & & 15.22$^{+1.86}_{-1.93}$ ($\times$10$^{20}$ cm$^{-2}$) \\
         $\chi^{2}$/dof & & 1207.87/1187 ($\simeq$1.02)\\
         \hline
      \end{tabular}
     }
\end{center}
\end{table}

The abundance ratios of \cite{Anders89} were used in
constructing Fig. \ref{fig:abund}. The number ratio relative to iron is then
$4.68\times 10^{-5}$, so the missing mass of
iron within 5 kpc, relative to unit central Solar abundance, is $1.4\times
10^6\Msun$ (i.e the iron required to increase the present observed
abundance from 0.37 to $1\Zsun$). 

While the two-temperature model is a much better fit than a single-temperature model for the X-ray emission in the cluster core, it is possible that the temperature structure is even more complex. An additional problem that affects the spectra of the cores of galaxy groups and clusters, is the temperature bias \citep{Buote01, Werner08}. This bias has the effect that, when complex structure in cluster cores is modelled using simple single-temperature models, the measured abundances can be biased towards lower values. For these reasons, although the RGS data indicate no third temperature component below 0.4 keV \citep{Sanders08}, we attempted to model the core emission using a three-temperature model, by adding an extra {\sc vapec} component to our two-temperature model. The normalisation and the temperature of this component were allowed to vary freely, while the abundances were tied as in the two-temperature model. We found that adding a third {\sc vapec} component did not significantly improve the fit to the data; the statistical improvement to the fit was relatively small ($\chi^{2}$/dof = 1198.89/1183). In addition, it was not possible to constrain the value of the temperature of the third component, and the lower limit on its normalisation was consistent with zero. However, we note that, even with the three-temperature model, the measured iron abundance remains low at 0.39 Z$\odot$. 

In addition to the two-temperature {\sc vapec} model, we fitted the spectra from the central 5 kpc with a combined single-temperature and cooling flow model, namely {\sc wabs*(vapec+vmcflow)}. In a cooling flow model \citep[for a review, see][]{Fabian94}, the gas in galaxy groups and clusters is assumed to be radiatively cooling from an upper temperature to a lower one. For this spectral model, the high-temperature value of the {\sc vmcflow} component was tied to the temperature of the {\sc vapec} component, and the abundances of the two components were tied. The Na abundance of the {\sc vmcflow} component was tied to that of Fe, and the value of its low-temperature component was allowed to vary freely. To prevent the attenuation of unphysically low values, the Ne abundance was tied to that of Fe, while the rest of the abundances were tied as in the {\sc wabs*(vapec+vapec)} model. This model was only a slightly worse fit statistically to the data than the two-temperature model, giving $\chi^{2}$/dof = 1219.59/1188 $\simeq$ 1.03. The measured Fe abundance remains low at 0.44 Z$\odot$, while the high- and low-temperature components were found to have values equal to 1.12 keV and 0.38 keV for the innermost bin, and 1.43 keV and 0.31 keV for the second innermost bin, respectively. 

As a final check, we show the radial abundance profiles of silicon and sulphur (middle and bottom panels of Fig. \ref{fig:abund}, respectively), which are less susceptible to the effect of low-temperature components. The profiles of silicon and sulphur are similar to that of iron. We conclude that the abundance drops observed towards the centre of NGC~4696 are real. We also attempted to determine the radial abundance profiles of other elements, such as argon. Unfortunately, the abundance profiles were highly variable, and could not be physically interpreted. 

The two-temperature model of the extended emission from the central 5 kpc of NGC~4696 is a significant improvement over the single-temperature model, and a three-temperature model does not improve the spectral fit. Therefore, our results obtained using a two-temperature model for the core emission are relatively unaffected by projection effects, and the aforementioned temperature bias.

\section{Where is the missing iron?}

We now assume that the observed iron abundance drop at the centre of the
Centaurus cluster is due to a real absence of iron, and explore where it
might be, or has gone. In this section, we compare the missing iron mass with the masses of hot and cold gas, and the dust within the
inner 5\,kpc of NGC\,4696.



A radius of 5\,kpc encompasses most of the optical filamentation seen
in the galaxy \citep{Fabian82, Sparks89, Crawford05b, Farage10, Canning11b}, as well as the dust
emission \citep{Kaneda05, Kaneda07, Mittal11} and far-infrared
emission from gas \citep{Mittal11}. The region is clearly multiphase with
gas and dust at less than 30\,K, warm gas at $10^3-10^4\K$, as well as the
X-ray emitting gas around $10^7\K$. The H$\alpha$ luminosity is about
$2\times 10^{40}\ergps$ \citep{Farage10, Canning11b}, which
implies a mass of warm gas of $4\times 10^5\Msun$ \citep{Farage10}. The mass of cold gas estimated through its CII emission is
$0.5-5\times 10^7\Msun$, and the mass of cold dust at
about 19\,K is $1.6\pm 0.3\times 10^6\Msun$ \citep{Mittal11}. 

Iron is usually depleted onto grains in cold gas \citep[see e.g.][]{Draine09}, and if we assume Solar abundance for that gas, then iron will be
about 20 per cent of its mass, or about $3.2\pm0.6\times 10^5\Msun$ of
iron in grains. The iron abundance of the cold gas could be 
higher than for the hot
gas, so a significant fraction of the missing iron could be in the
grains \citep[see also][]{Canning11a}. Feedback activity caused by the nucleus of the galaxy is
demonstrated by the bubbles of radio-emitting plasma evident in
Fig. \ref{fig:expcorr}, which also lie within the inner 5\,kpc. The bubbles are
buoyant in the intracluster gas, and will rise, dragging the
filamentation outward, as discussed for these and other cluster
filaments \citep{Fabian03, Crawford05b}. This means
that grains are transported outward by the quasi-continuous bubbling
process necessary to prevent a full cooling flow developing in the
cluster core \citep[for a review, see][]{Fabian12}. If filaments and grains
are somehow destroyed beyond 10--20\,kpc, e.g. by sputtering in hot gas, then iron will be returned
to the hot medium there, contributing to its high iron abundance
(see the top panel of Fig. \ref{fig:abund}). Silicon can also be depleted onto grains, and possibly sulphur too \citep[see section 2.3 in][and references therein]{Calura09}, so explaining their low central abundance values. 

An alternative way a central abundance drop can arise, is if the iron injected by Type Ia SNe remains clumped, rather than mixing with the surrounding gas. High metallicity gas clumps cool radiatively on a relatively short timescale, giving much weaker overall iron emission. This is because, while in a high metallicity gas clump the iron ions only have to cool themselves, if they are dispersed in a low metallicity gas cloud, they have to cool the surrounding hydrogen, helium etc, as well as themselves. This requires the iron emission to be stronger \citep[see][]{Fabian01, Morris03}. A problem that arises when trying to apply this solution to the Centaurus cluster is that the abundance drop would not be expected to coincide so well with the dust extent; rather, it would lead to a more extended drop. Nevertheless, this mechanism could play a lesser role in shaping the overall iron abundance profile.  

The inner multiphase medium of NGC\,4696 is dominated by stellar mass
loss, including ejecta from supernovae, and has been modelled by
\cite{Graham06} and \cite{Sanders06a}. The fate of stellar
mass loss in central cluster galaxies has been further emphasised by
\cite{Voit11}. 
Stellar mass loss and  Type Ia supernovae (SNIa) are the sources of
iron in the inner region of central cluster galaxies. We estimate the
injection of iron from Type Ia SN within the inner 5\,kpc to be about
$3.4\times10^{-3}\Msunpyr$ of iron, using the SN rate given in
\cite{Pellegrini06}, a total $B$-band luminosity of $10^{11}\Lsun$
\citep{Sanders06a} and the Hernquist stellar profile of \cite{Farage10}, according to which 43\% of the projected light is contained within a 5 kpc radius. This means that the missing iron corresponds to about
0.4\,Gyr of SN injection within 5\,kpc. For stellar mass loss, we obtain
an injection rate of $1.3\Msunpyr$ for within 5\,kpc,
and an iron injection rate (assumed to be at
Solar abundance) of $3.4\times 10^{-3}\Msunpyr$, identical to the SNIa
rate, using the \cite{Pellegrini06} rate used
by \cite{Sanders06a}, assuming an age of $10^{10}\yr$. Much of
that iron can have originated in Type II SN during the early growth of
the galaxy. From abundance ratios, \cite{Sanders06a} find that 
the main metallicity peak has been enriched 70 per cent by SNIa and 30
per cent by SNII.  

\cite{Voit11} suggest that some of the stellar mass
loss remains cool and distinct from the surrounding hot gas, and thus
preserves its embedded dust, making the cold filaments
dusty. Ultraviolet observations of the stripped wake of the giant star Mira
in our galaxy support this hypothesis \citep{Martin07}. It is thus
possible that much of the iron in the stellar mass loss remains in
grain form, and is incorporated into the filaments. The only hot atomic
iron in the inner region is then that due to SNIa. 

The particle heating model of \cite{Ferland09} successfully
explains the line ratios of cluster filament systems, such as that
around NGC\,4696. The surrounding hot gas at temperature $10^7T_7\K$
interpenetrates the filaments, and becomes incorporated in
them at a rate of $\sim 0.7 L({\rm H}\alpha)_{40} T_7^{-1}\Msunpyr$,
producing an H$\alpha$ luminosity of $10^{40}L({\rm H}\alpha)_{40}\ergps$.  
For NGC\,4696, this means a rate of about $1.4\Msunpyr$, where T$_{7}\sim$1, so that most of
the inner gas may be processed in this way in $\sim 5\times 10^8\yr$.   

We note that, if the SNIa-injected iron remains localised, then high iron
abundance clumps of hot gas will result. Such clumps can perhaps explain the detection of [Fe {\sc x}]${\rm \lambda}$6374 coronal line emission from iron \citep{Canning11a}. These clumps can cool rapidly,
triggering iron-rich cold clouds \citep{Fabian01, Morris03}. The cold iron can then grow onto grains.

The low measured oxygen abundance in the central 5 kpc (see Table \ref{tab:results}) is explained if we assume that most oxygen is also in cold dust grains. Type Ia SN produce a negligible amount of oxygen, compared to iron, so that the hot gas naturally has a low central oxygen abundance. This result can also help explain the weakness of O {\sc vii} emission seen in cool core clusters \citep{Sanders11}. The appearance, abundances and quantities of gas at temperatures of 10$^{6}$--10$^{7}$ K in this environment are very complicated. 

We assume, then, that there is a quasi-continuous cyclical process of
bubbling, which leads to the dusty clouds being eventually dragged away 
from the centre. The age of the present bubbles is $\sim 6\times
10^6\yr$ \citep{Dunn06}, and the  time for each cycle could be slightly longer,
at $10^7\yr$. At the above present SNII iron injection rate, it takes $9.4\times
10^7\yr$ to accumulate the level of iron inferred from the observed
mass of dust. Thus, an approximate steady state is reached if about 10
per cent of the filaments, with their embedded  grains, is dragged out
from the central 5\,kpc in each cycle. Some of the SNIa enriched hot gas
will also be dragged outward.
 
\section{Discussion}

The drop in iron abundance observed within the central few kpc of the
Centaurus cluster is plausibly due to the iron injected by stellar
mass loss remaining bound in dust grains, as proposed by \cite{Voit11}. The missing iron is then in the grains, seen at optical
and far-infrared wavelengths in the dusty filaments occupying the same
spatial region as the abundance drop, and at larger radii in both cold
and hot gas where it has been transported by the bubbling process of 
feedback in the cluster core. Grains provide a plausible method for separating iron and other refractory metals from hydrogen, so leading to large central abundance drops. 

Further work on the scenario outlined here requires further observations of
the detailed metallicity of the stars and cold gas in NGC\,4696, as
well as of the hot gas. Deeper Chandra observations, in particular, are needed. The inner region of NGC~4696, at the centre of the Centaurus cluster, is a unique nearby environment for studying stellar mass loss in giant elliptical galaxies.

\section*{Acknowledgements}
We thank Raymond Oonk for helpful discussions. EKP acknowledges the support of a STFC sudentship. ACF thanks the Royal Society for support. We would like to thank the anonymous referee for helpful comments, that improved the content of this paper.  

The plots in this paper were created using {\sc veusz}.\footnote{http://home.gna.org/veusz/}

\bibliographystyle{mn2e}
\bibliography{references}

\end{document}

%% file: defn.tex
\newcommand{\yr}{\rm\thinspace yr}

\newcommand{\s}{\rm\thinspace s}

\newcommand{\K}{\rm\thinspace K}

\newcommand{\Msun}{\hbox{$\rm\thinspace M_{\odot}$}}

\newcommand{\Msunpyr}{\hbox{$\Msun\yr^{-1}\,$}}

\newcommand{\erg}{\rm\thinspace erg}

\newcommand{\ergps}{\hbox{$\erg\s^{-1}\,$}}

\newcommand{\Lsun}{\hbox{$\rm\thinspace L_{\odot}$}}

\newcommand{\Zsun}{\hbox{$\thinspace \mathrm{Z}_{\odot}$}}














%% file: cen_acf.bbl
\begin{thebibliography}{}

\bibitem[\protect\citeauthoryear{{Allen}}{{Allen}}{1995}]{Allen95}
{Allen} S.~W.,  1995, \mnras, 276, 947

\bibitem[\protect\citeauthoryear{{Allen} \& {Fabian}}{{Allen} \&
  {Fabian}}{1994}]{Allen94}
{Allen} S.~W.,  {Fabian} A.~C.,  1994, \mnras, 269, 409

\bibitem[\protect\citeauthoryear{{Allen}, {Fabian}, {Johnstone}, {Arnaud} \&
  {Nulsen}}{{Allen} et~al.}{2001}]{Allen01}
{Allen} S.~W.,  {Fabian} A.~C.,  {Johnstone} R.~M.,  {Arnaud} K.~A.,
  {Nulsen} P.~E.~J.,  2001, \mnras, 322, 589

\bibitem[\protect\citeauthoryear{{Anders} \& {Grevesse}}{{Anders} \&
  {Grevesse}}{1989}]{Anders89}
{Anders} E.,  {Grevesse} N.,  1989, \gca, 53, 197

\bibitem[\protect\citeauthoryear{{Arnaud}}{{Arnaud}}{1996}]{Arnaud96}
{Arnaud} K.~A.,  1996, in {Jacoby} G.~H.,  {Barnes} J.,  eds, Astronomical Data
  Analysis Software and Systems V Vol.~101 of Astronomical Society of the
  Pacific Conference Series, {XSPEC: The First Ten Years}.
p.~17

\bibitem[\protect\citeauthoryear{{B{\^i}rzan}, {Rafferty}, {McNamara}, {Wise}
  \& {Nulsen}}{{B{\^i}rzan} et~al.}{2004}]{Birzan04}
{B{\^i}rzan} L.,  {Rafferty} D.~A.,  {McNamara} B.~R.,  {Wise} M.~W.,
  {Nulsen} P.~E.~J.,  2004, \apj, 607, 800

\bibitem[\protect\citeauthoryear{{Buote}}{{Buote}}{2001}]{Buote01}
{Buote} D.~A.,  2001, \apj, 548, 652

\bibitem[\protect\citeauthoryear{{Calura}, {Dessauges-Zavadski}, {Prochaska} \&
  {Matteucci}}{{Calura} et~al.}{2009}]{Calura09}
{Calura} F.,  {Dessauges-Zavadski} M.,  {Prochaska} J.~X.,    {Matteucci} F.,
  2009, \apj, 693, 1236

\bibitem[\protect\citeauthoryear{{Canning}, {Fabian}, {Johnstone}, {Sanders},
  {Crawford}, {Ferland} \& {Hatch}}{{Canning} et~al.}{2011}]{Canning11b}
{Canning} R.~E.~A.,  {Fabian} A.~C.,  {Johnstone} R.~M.,  {Sanders} J.~S.,
  {Crawford} C.~S.,  {Ferland} G.~J.,    {Hatch} N.~A.,  2011, \mnras, 417,
  3080

\bibitem[\protect\citeauthoryear{{Canning}, {Fabian}, {Johnstone}, {Sanders},
  {Crawford}, {Hatch} \& {Ferland}}{{Canning} et~al.}{2011}]{Canning11a}
{Canning} R.~E.~A.,  {Fabian} A.~C.,  {Johnstone} R.~M.,  {Sanders} J.~S.,
  {Crawford} C.~S.,  {Hatch} N.~A.,    {Ferland} G.~J.,  2011, \mnras, 411, 411

\bibitem[\protect\citeauthoryear{{Churazov}, {Forman}, {Jones} \&
  {B{\"o}hringer}}{{Churazov} et~al.}{2003}]{Churazov03}
{Churazov} E.,  {Forman} W.,  {Jones} C.,    {B{\"o}hringer} H.,  2003, \apj,
  590, 225

\bibitem[\protect\citeauthoryear{{Crawford}, {Hatch}, {Fabian} \&
  {Sanders}}{{Crawford} et~al.}{2005}]{Crawford05b}
{Crawford} C.~S.,  {Hatch} N.~A.,  {Fabian} A.~C.,    {Sanders} J.~S.,  2005,
  \mnras, 363, 216

\bibitem[\protect\citeauthoryear{{de Jong}, {Norgaard-Nielsen}, {Jorgensen} \&
  {Hansen}}{{de Jong} et~al.}{1990}]{deJong90}
{de Jong} T.,  {Norgaard-Nielsen} H.~U.,  {Jorgensen} H.~E.,    {Hansen} L.,
  1990, \aap, 232, 317

\bibitem[\protect\citeauthoryear{{de Plaa}, {Werner}, {Bleeker}, {Vink},
  {Kaastra} \& {M{\'e}ndez}}{{de Plaa} et~al.}{2007}]{dePlaa07}
{de Plaa} J.,  {Werner} N.,  {Bleeker} J.~A.~M.,  {Vink} J.,  {Kaastra} J.~S.,
    {M{\'e}ndez} M.,  2007, \aap, 465, 345

\bibitem[\protect\citeauthoryear{{Draine}}{{Draine}}{2009}]{Draine09}
{Draine} B.~T.,  2009, in {Henning} T.,  {Gr{\"u}n} E.,   {Steinacker} J.,
  eds, Cosmic Dust - Near and Far Vol.~414 of Astronomical Society of the
  Pacific Conference Series, {Interstellar Dust Models and Evolutionary
  Implications}.
p.~453

\bibitem[\protect\citeauthoryear{{Dunn} \& {Fabian}}{{Dunn} \&
  {Fabian}}{2006}]{Dunn06}
{Dunn} R.~J.~H.,  {Fabian} A.~C.,  2006, \mnras, 373, 959

\bibitem[\protect\citeauthoryear{{Edge}}{{Edge}}{2001}]{Edge01}
{Edge} A.~C.,  2001, \mnras, 328, 762

\bibitem[\protect\citeauthoryear{{Edge} \& {Stewart}}{{Edge} \&
  {Stewart}}{1991}]{Edge91}
{Edge} A.~C.,  {Stewart} G.~C.,  1991, \mnras, 252, 414

\bibitem[\protect\citeauthoryear{{Fabian}}{{Fabian}}{1994}]{Fabian94}
{Fabian} A.~C.,  1994, \araa, 32, 277

\bibitem[\protect\citeauthoryear{{Fabian}}{{Fabian}}{2012}]{Fabian12}
{Fabian} A.~C.,  2012, \araa, 50, 455

\bibitem[\protect\citeauthoryear{{Fabian}, {Mushotzky}, {Nulsen} \&
  {Peterson}}{{Fabian} et~al.}{2001}]{Fabian01}
{Fabian} A.~C.,  {Mushotzky} R.~F.,  {Nulsen} P.~E.~J.,    {Peterson} J.~R.,
  2001, \mnras, 321, L20

\bibitem[\protect\citeauthoryear{{Fabian}, {Nulsen}, {Atherton} \&
  {Taylor}}{{Fabian} et~al.}{1982}]{Fabian82}
{Fabian} A.~C.,  {Nulsen} P.~E.~J.,  {Atherton} P.~D.,    {Taylor} K.,  1982,
  \mnras, 201, 17P

\bibitem[\protect\citeauthoryear{{Fabian}, {Sanders}, {Crawford}, {Conselice},
  {Gallagher} \& {Wyse}}{{Fabian} et~al.}{2003}]{Fabian03}
{Fabian} A.~C.,  {Sanders} J.~S.,  {Crawford} C.~S.,  {Conselice} C.~J.,
  {Gallagher} J.~S.,    {Wyse} R.~F.~G.,  2003, \mnras, 344, L48

\bibitem[\protect\citeauthoryear{{Fabian}, {Sanders}, {Taylor} \&
  {Allen}}{{Fabian} et~al.}{2005}]{Fabian05}
{Fabian} A.~C.,  {Sanders} J.~S.,  {Taylor} G.~B.,    {Allen} S.~W.,  2005,
  \mnras, 360, L20

\bibitem[\protect\citeauthoryear{{Farage}, {McGregor}, {Dopita} \&
  {Bicknell}}{{Farage} et~al.}{2010}]{Farage10}
{Farage} C.~L.,  {McGregor} P.~J.,  {Dopita} M.~A.,    {Bicknell} G.~V.,  2010,
  \apj, 724, 267

\bibitem[\protect\citeauthoryear{{Ferland}, {Fabian}, {Hatch}, {Johnstone},
  {Porter}, {van Hoof} \& {Williams}}{{Ferland} et~al.}{2009}]{Ferland09}
{Ferland} G.~J.,  {Fabian} A.~C.,  {Hatch} N.~A.,  {Johnstone} R.~M.,  {Porter}
  R.~L.,  {van Hoof} P.~A.~M.,    {Williams} R.~J.~R.,  2009, \mnras, 392, 1475

\bibitem[\protect\citeauthoryear{{Finoguenov}, {David} \&
  {Ponman}}{{Finoguenov} et~al.}{2000}]{Finoguenov00}
{Finoguenov} A.,  {David} L.~P.,    {Ponman} T.~J.,  2000, \apj, 544, 188

\bibitem[\protect\citeauthoryear{{Fukazawa}, {Ohashi}, {Fabian}, {Canizares},
  {Ikebe}, {Makishima}, {Mushotzky} \& {Yamashita}}{{Fukazawa}
  et~al.}{1994}]{Fukazawa94}
{Fukazawa} Y.,  {Ohashi} T.,  {Fabian} A.~C.,  {Canizares} C.~R.,  {Ikebe} Y.,
  {Makishima} K.,  {Mushotzky} R.~F.,    {Yamashita} K.,  1994, \pasj, 46, L55

\bibitem[\protect\citeauthoryear{{Gilfanov}, {Syunyaev} \&
  {Churazov}}{{Gilfanov} et~al.}{1987}]{Gilfanov87}
{Gilfanov} M.~R.,  {Syunyaev} R.~A.,    {Churazov} E.~M.,  1987, Soviet
  Astronomy Letters, 13, 3

\bibitem[\protect\citeauthoryear{{Graham}, {Fabian}, {Sanders} \&
  {Morris}}{{Graham} et~al.}{2006}]{Graham06}
{Graham} J.,  {Fabian} A.~C.,  {Sanders} J.~S.,    {Morris} R.~G.,  2006,
  \mnras, 368, 1369

\bibitem[\protect\citeauthoryear{{Johnstone}, {Allen}, {Fabian} \&
  {Sanders}}{{Johnstone} et~al.}{2002}]{Johnstone02}
{Johnstone} R.~M.,  {Allen} S.~W.,  {Fabian} A.~C.,    {Sanders} J.~S.,  2002,
  \mnras, 336, 299

\bibitem[\protect\citeauthoryear{{Johnstone}, {Fabian} \& {Nulsen}}{{Johnstone}
  et~al.}{1987}]{Johnstone87}
{Johnstone} R.~M.,  {Fabian} A.~C.,    {Nulsen} P.~E.~J.,  1987, \mnras, 224,
  75

\bibitem[\protect\citeauthoryear{{Johnstone}, {Hatch}, {Ferland}, {Fabian},
  {Crawford} \& {Wilman}}{{Johnstone} et~al.}{2007}]{Johnstone07}
{Johnstone} R.~M.,  {Hatch} N.~A.,  {Ferland} G.~J.,  {Fabian} A.~C.,
  {Crawford} C.~S.,    {Wilman} R.~J.,  2007, \mnras, 382, 1246

\bibitem[\protect\citeauthoryear{{Kaneda}, {Onaka}, {Kitayama}, {Okada} \&
  {Sakon}}{{Kaneda} et~al.}{2007}]{Kaneda07}
{Kaneda} H.,  {Onaka} T.,  {Kitayama} T.,  {Okada} Y.,    {Sakon} I.,  2007,
  \pasj, 59, 107

\bibitem[\protect\citeauthoryear{{Kaneda}, {Onaka} \& {Sakon}}{{Kaneda}
  et~al.}{2005}]{Kaneda05}
{Kaneda} H.,  {Onaka} T.,    {Sakon} I.,  2005, \apjl, 632, L83

\bibitem[\protect\citeauthoryear{{Kirkpatrick}, {McNamara}, {Rafferty},
  {Nulsen}, {B{\^i}rzan}, {Kazemzadeh}, {Wise}, {Gitti} \&
  {Cavagnolo}}{{Kirkpatrick} et~al.}{2009}]{Kirkpatrick09}
{Kirkpatrick} C.~C.,  {McNamara} B.~R.,  {Rafferty} D.~A.,  {Nulsen} P.~E.~J.,
  {B{\^i}rzan} L.,  {Kazemzadeh} F.,  {Wise} M.~W.,  {Gitti} M.,    {Cavagnolo}
  K.~W.,  2009, \apj, 697, 867

\bibitem[\protect\citeauthoryear{{Loewenstein}}{{Loewenstein}}{2006}]{Loewenstein06}
{Loewenstein} M.,  2006, \apj, 648, 230

\bibitem[\protect\citeauthoryear{{Lucey}, {Currie} \& {Dickens}}{{Lucey}
  et~al.}{1986}]{Lucey86}
{Lucey} J.~R.,  {Currie} M.~J.,    {Dickens} R.~J.,  1986, \mnras, 221, 453

\bibitem[\protect\citeauthoryear{{Martin}, {Seibert}, {Neill}, {Schiminovich},
  {Forster}, {Rich}, {Welsh}, {Madore}, {Wheatley}, {Morrissey} \&
  {Barlow}}{{Martin} et~al.}{2007}]{Martin07}
{Martin} D.~C.,  {Seibert} M.,  {Neill} J.~D.,  {Schiminovich} D.,  {Forster}
  K.,  {Rich} R.~M.,  {Welsh} B.~Y.,  {Madore} B.~F.,  {Wheatley} J.~M.,
  {Morrissey} P.,    {Barlow} T.~A.,  2007, \nat, 448, 780

\bibitem[\protect\citeauthoryear{{McNamara} \& {Nulsen}}{{McNamara} \&
  {Nulsen}}{2007}]{McNamara07}
{McNamara} B.~R.,  {Nulsen} P.~E.~J.,  2007, \araa, 45, 117

\bibitem[\protect\citeauthoryear{{McNamara} \& {Nulsen}}{{McNamara} \&
  {Nulsen}}{2012}]{McNamara12}
{McNamara} B.~R.,  {Nulsen} P.~E.~J.,  2012, New Journal of Physics, 14, 055023

\bibitem[\protect\citeauthoryear{{Mittal}, {O'Dea}, {Ferland}, {Oonk}, {Edge},
  {Canning}, {Russell}, {Baum}, {B{\"o}hringer}, {Combes}, {Donahue}, {Fabian},
  {Hatch}, {Hoffer}, {Johnstone}, {McNamara}, {Salom{\'e}} \&
  {Tremblay}}{{Mittal} et~al.}{2011}]{Mittal11}
{Mittal} R.,  {O'Dea} C.~P.,  {Ferland} G.,  {Oonk} J.~B.~R.,  {Edge} A.~C.,
  {Canning} R.~E.~A.,  {Russell} H.,  {Baum} S.~A.,  {B{\"o}hringer} H.,
  {Combes} F.,  {Donahue} M.,  {Fabian} A.~C.,  {Hatch} N.~A.,  {Hoffer} A.,
  {Johnstone} R.,  {McNamara} B.~R.,  {Salom{\'e}} P.,    {Tremblay} G.,  2011,
  \mnras, 418, 2386

\bibitem[\protect\citeauthoryear{{Molendi}, {De Grandi} \&
  {Guainazzi}}{{Molendi} et~al.}{2002}]{Molendi02}
{Molendi} S.,  {De Grandi} S.,    {Guainazzi} M.,  2002, \aap, 392, 13

\bibitem[\protect\citeauthoryear{{Morris} \& {Fabian}}{{Morris} \&
  {Fabian}}{2003}]{Morris03}
{Morris} R.~G.,  {Fabian} A.~C.,  2003, \mnras, 338, 824

\bibitem[\protect\citeauthoryear{{Nulsen}, {Johnstone} \& {Fabian}}{{Nulsen}
  et~al.}{1987}]{Nulsen87}
{Nulsen} P.~E.~J.,  {Johnstone} R.~M.,    {Fabian} A.~C.,  1987, Proceedings of
  the Astronomical Society of Australia, 7, 132

\bibitem[\protect\citeauthoryear{{O'Dea}, {Baum}, {Privon}, {Noel-Storr},
  {Quillen}, {Zufelt}, {Park}, {Edge}, {Russell}, {Fabian}, {Donahue},
  {Sarazin}, {McNamara}, {Bregman} \& {Egami}}{{O'Dea} et~al.}{2008}]{Odea08}
{O'Dea} C.~P.,  {Baum} S.~A.,  {Privon} G.,  {Noel-Storr} J.,  {Quillen} A.~C.,
   {Zufelt} N.,  {Park} J.,  {Edge} A.,  {Russell} H.,  {Fabian} A.~C.,
  {Donahue} M.,  {Sarazin} C.~L.,  {McNamara} B.,  {Bregman} J.~N.,    {Egami}
  E.,  2008, \apj, 681, 1035

\bibitem[\protect\citeauthoryear{{O'Dea}, {Quillen}, {O'Dea}, {Tremblay},
  {Snios}, {Baum}, {Christiansen}, {Noel-Storr}, {Edge}, {Donahue} \&
  {Voit}}{{O'Dea} et~al.}{2010}]{Odea10}
{O'Dea} K.~P.,  {Quillen} A.~C.,  {O'Dea} C.~P.,  {Tremblay} G.~R.,  {Snios}
  B.~T.,  {Baum} S.~A.,  {Christiansen} K.,  {Noel-Storr} J.,  {Edge} A.~C.,
  {Donahue} M.,    {Voit} G.~M.,  2010, \apj, 719, 1619

\bibitem[\protect\citeauthoryear{{Pellegrini} \& {Ciotti}}{{Pellegrini} \&
  {Ciotti}}{2006}]{Pellegrini06}
{Pellegrini} S.,  {Ciotti} L.,  2006, \mnras, 370, 1797

\bibitem[\protect\citeauthoryear{{Peterson} \& {Fabian}}{{Peterson} \&
  {Fabian}}{2006}]{Peterson06}
{Peterson} J.~R.,  {Fabian} A.~C.,  2006, \physrep, 427, 1

\bibitem[\protect\citeauthoryear{{Rafferty}, {B{\^i}rzan}, {Nulsen},
  {McNamara}, {Brandt}, {Wise} \& {R{\"o}ttgering}}{{Rafferty}
  et~al.}{2013}]{Rafferty13}
{Rafferty} D.~A.,  {B{\^i}rzan} L.,  {Nulsen} P.~E.~J.,  {McNamara} B.~R.,
  {Brandt} W.~N.,  {Wise} M.~W.,    {R{\"o}ttgering} H.~J.~A.,  2013, \mnras,
  428, 58

\bibitem[\protect\citeauthoryear{{Rafferty}, {McNamara}, {Nulsen} \&
  {Wise}}{{Rafferty} et~al.}{2006}]{Rafferty06}
{Rafferty} D.~A.,  {McNamara} B.~R.,  {Nulsen} P.~E.~J.,    {Wise} M.~W.,
  2006, \apj, 652, 216

\bibitem[\protect\citeauthoryear{{Rasmussen} \& {Ponman}}{{Rasmussen} \&
  {Ponman}}{2007}]{Rasmussen07}
{Rasmussen} J.,  {Ponman} T.~J.,  2007, \mnras, 380, 1554

\bibitem[\protect\citeauthoryear{{Russell}, {Sanders} \& {Fabian}}{{Russell}
  et~al.}{2008}]{Russell08}
{Russell} H.~R.,  {Sanders} J.~S.,    {Fabian} A.~C.,  2008, \mnras, 390, 1207

\bibitem[\protect\citeauthoryear{{Sanders} \& {Fabian}}{{Sanders} \&
  {Fabian}}{2002}]{Sanders02}
{Sanders} J.~S.,  {Fabian} A.~C.,  2002, \mnras, 331, 273

\bibitem[\protect\citeauthoryear{{Sanders} \& {Fabian}}{{Sanders} \&
  {Fabian}}{2006a}]{Sanders06a}
{Sanders} J.~S.,  {Fabian} A.~C.,  2006a, \mnras, 371, 1483

\bibitem[\protect\citeauthoryear{{Sanders} \& {Fabian}}{{Sanders} \&
  {Fabian}}{2006b}]{Sanders06b}
{Sanders} J.~S.,  {Fabian} A.~C.,  2006b, \mnras, 370, 63

\bibitem[\protect\citeauthoryear{{Sanders} \& {Fabian}}{{Sanders} \&
  {Fabian}}{2007}]{Sanders07}
{Sanders} J.~S.,  {Fabian} A.~C.,  2007, \mnras, 381, 1381

\bibitem[\protect\citeauthoryear{{Sanders} \& {Fabian}}{{Sanders} \&
  {Fabian}}{2011}]{Sanders11}
{Sanders} J.~S.,  {Fabian} A.~C.,  2011, \mnras, 412, L35

\bibitem[\protect\citeauthoryear{{Sanders}, {Fabian}, {Allen}, {Morris},
  {Graham} \& {Johnstone}}{{Sanders} et~al.}{2008}]{Sanders08}
{Sanders} J.~S.,  {Fabian} A.~C.,  {Allen} S.~W.,  {Morris} R.~G.,  {Graham}
  J.,    {Johnstone} R.~M.,  2008, \mnras, 385, 1186

\bibitem[\protect\citeauthoryear{{Sanders}, {Fabian}, {Allen} \&
  {Schmidt}}{{Sanders} et~al.}{2004}]{Sanders04}
{Sanders} J.~S.,  {Fabian} A.~C.,  {Allen} S.~W.,    {Schmidt} R.~W.,  2004,
  \mnras, 349, 952

\bibitem[\protect\citeauthoryear{{Schmidt}, {Fabian} \& {Sanders}}{{Schmidt}
  et~al.}{2002}]{Schmidt02}
{Schmidt} R.~W.,  {Fabian} A.~C.,    {Sanders} J.~S.,  2002, \mnras, 337, 71

\bibitem[\protect\citeauthoryear{{Sparks}, {Macchetto} \& {Golombek}}{{Sparks}
  et~al.}{1989}]{Sparks89}
{Sparks} W.~B.,  {Macchetto} F.,    {Golombek} D.,  1989, \apj, 345, 153

\bibitem[\protect\citeauthoryear{{Struble} \& {Rood}}{{Struble} \&
  {Rood}}{1999}]{Struble99}
{Struble} M.~F.,  {Rood} H.~J.,  1999, \apjs, 125, 35

\bibitem[\protect\citeauthoryear{{Voit} \& {Donahue}}{{Voit} \&
  {Donahue}}{2011}]{Voit11}
{Voit} G.~M.,  {Donahue} M.,  2011, \apjl, 738, L24

\bibitem[\protect\citeauthoryear{{Werner}, {Durret}, {Ohashi}, {Schindler} \&
  {Wiersma}}{{Werner} et~al.}{2008}]{Werner08}
{Werner} N.,  {Durret} F.,  {Ohashi} T.,  {Schindler} S.,    {Wiersma}
  R.~P.~C.,  2008, \ssr, 134, 337

\end{thebibliography}
